\documentclass[pdflatex,sn-mathphys-num]{sn-jnl}% Math and Physical Sciences Numbered Reference Style
\usepackage{graphicx}%
\usepackage{multirow}%
\usepackage{amsmath,amssymb,amsfonts}%
\usepackage{amsthm}%
\usepackage{mathrsfs}%
\usepackage[title]{appendix}%
\usepackage{xcolor}%
\usepackage{hyperref}
\usepackage{textcomp}%
\usepackage{manyfoot}%
\usepackage{booktabs}%
\usepackage{algorithm}%
\usepackage{algorithmicx}%
\usepackage{algpseudocode}%
\usepackage{listings}%
\usepackage{geometry}
\newcommand{\chihyb}{\chi^{(2)}{}_{\mathrm{eff}}}
\newcommand{\chiML}{\chi^{(2)}{}_{\mathrm{sub}}}
\newcommand{\chimol}{\chi^{(2)}{}_{\mathrm{mol}}}
\newcommand{\chiEF}{\chi^{(2)}{}_{\mathrm{EFISH}}}
\newcommand{\chithree}{\chi^{(3)}{}_{\mathrm{sub}}}

\begin{document}

\title{Rational Design of Low-Dimensional Hybrid Organic/Inorganic Interfaces for Enhanced Second-Harmonic Generation}

%%=============================================================%%
%% GivenName	-> \fnm{Joergen W.}
%% Particle	-> \spfx{van der} -> surname prefix
%% FamilyName	-> \sur{Ploeg}
%% Suffix	-> \sfx{IV}
%% \author*[1,2]{\fnm{Joergen W.} \spfx{van der} \sur{Ploeg} 
%%  \sfx{IV}}\email{iauthor@gmail.com}
%%=============================================================%%

\author[1]{\fnm{Michele} \sur{Guerrini}}\email{michele.guerrini@uni-jena.de}

\author[1]{\fnm{M. Sufyan} \sur{Ramzan}}\email{muhammad.ramzan@uni-jena.de}

\author[1,2]{\fnm{Caterina} \sur{Cocchi}}\email{caterina.cocchi@uni-jena.de}

\affil[1]{\orgdiv{Institut f\"ur Festk\"orpertheorie und -optik,}
\orgname{Friedrich-Schiller-Universit\"at Jena,}
\orgaddress{\state{07743 Jena,}
\country{Germany}}}
\affil[2]{\orgdiv{Abbe Center of Photonics,}
\orgname{Friedrich-Schiller-Universit\"at Jena,}
\orgaddress{\state{07745 Jena,}
\country{Germany}}}

%%==================================%%
%% Sample for unstructured abstract %%
%%==================================%%

\abstract{Hybrid interfaces formed by push-pull organic molecules physisorbed on two-dimensional (2D) semiconductors provide a structurally tunable platform for engineering second-harmonic generation (SHG). However, predicting and optimizing their macroscopic response remains a formidable challenge due to the phase-sensitive interference between the nonlinear responses of the adlayer and substrate. Here, we develop a physics-informed computational screening framework that decomposes the effective second-order susceptibility into its constituent substrate, molecular, and electric-field-induced SHG channels. Validated by fully atomistic first-principles calculations across representative interface structures, this model maps the complete orientation-resolved tensor landscape of chemically modulated carbon-conjugated polar molecules on 2D substrates with distinct symmetry. Crucially, our findings dismantle the conventional reliance on static polar descriptors, demonstrating that the ground-state permanent dipole magnitude is a fundamentally unreliable proxy for macroscopic nonlinear response. Instead, we show that the ultimate criterion for SHG maximization is encoded in the phase-resolved projection of the full anisotropic molecular hyperpolarizability tensor. On non-centrosymmetric substrates, the coherent interference between the physisorbed adlayer and the underlying 2D matrix induces a characteristic Fano-like asymmetry and ranking reversals that are entirely absent on centrosymmetric platforms. By establishing that functionalization topology, dynamic spatial orientation, and substrate point-group symmetry constitute a single, non-separable parameter space, this work provides a systematic blueprint for the rational assembly, predictive discovery, and non-invasive characterization of next-generation low-dimensional hybrid materials for nonlinear optoelectronics.}


\maketitle
\newpage

%% ============================================================
\section{Introduction}
\label{sec:intro}
%% ============================================================
Hybrid interfaces formed by two-dimensional (2D) semiconductors and physisorbed carbon-conjugated molecules are emerging as a premier platform for nanoscale optoelectronics~\cite{nied+21es,ji+22ns,zhan+24csr,cui+25cr}. Although extensive efforts have focused so far on electronic band alignment~\cite{koch21apl,chen+21jpcl,krum+21es,mela+22pccp,guo+22nr}, excitonic properties and dynamics~\cite{gonz+22prm,mark+22nano,schw+22jcp,tand+24pssa,xion+25nano}, and charge transfer~\cite{chen+20nano,jaco+22acsanm,vale+26rscai}, the potential of these interfaces as platforms for nonlinear optics remains largely unexplored. In particular, second-harmonic generation (SHG), with its exquisite sensitivity to symmetry and local environment~\cite{herr+25natph,bao+26prb,guo+26arxiv,patr-cocc26arxiv}, is particularly appealing as a probe for interfacial engineering~\cite{chen+26natcom}. 

A physisorbed organic molecule reshapes the SHG response of a 2D substrate in two complementary ways. It modifies the local interfacial electronic structure through frontier-orbital alignment, potentially introducing localized molecular states within the substrate gap, and, most importantly for SHG, it locally breaks substrate symmetry through its directional adsorption geometry, activating nonlinear optical channels that are weak or forbidden in the bare substrate~\cite{chen+26natcom,Autere2018}. Both effects are chemically tunable through edge functionalization of the adsorbates. Electron-donating or electron-withdrawing substituents, which modify frontier-orbital energies, may introduce a permanent dipole moment~\cite{Weil2010} and overall affect the optical response both in the linear and in the non-linear regime~\cite{andr-meat93jpb,zhao+22cs}.

Since SHG vanishes in centrosymmetric systems, its polarization-resolved response elegantly encodes the full tensorial structure of the interfacial second-order susceptibility~\cite{bao+26prb,guo+26arxiv,patr-cocc26arxiv,egem-ohno24jcp}. However, fully atomistic SHG simulations of realistic materials~\cite{atta+19prm,pike-patc21jpcc,ruan+24nl}, including hybrid interfaces~\cite{ma+23jpcc,chen+24jpcc}, remain computationally prohibitive due to the intense demands of evaluating high-order perturbation theory~\cite{prus+23epjst} on large, chemically complex cells. While a recent development of the polarizable continuum model for layered substrates (\texttt{LayerPCM})~\cite{Krumland2021b} in connection with \textit{ab initio} calculations allows an accurate description of the electronic structure of complex adsorbate geometries~\cite{tand+24pssa,Krumland2024a,krum+24pssa}, the community still misses an explicit physical framework that translates the nonlinear optical properties of isolated building-block calculations into a predictive model for hybrid interfaces.

In this work, we bridge this gap by developing a physics-informed screening framework that analytically reconstructs the total interfacial second-order susceptibility, elaborating with symmetry arguments the individual response of the constituents. Our model decomposes the total signal into substrate, molecule, and dipole-induced contributions. Applying this framework to asymmetrically fluorinated perylene derivatives, taken here as a prototypical class of polar dyes, and simulating their adsorption on substrates belonging to three different symmetry groups, we demonstrate that the heuristic assumption that the permanent dipole magnitude dictates SHG enhancement is not generally valid. Instead, we show that the decisive factor is the orientation-dependent projection of the full molecular hyperpolarizability tensor onto the probed polarization channels. This leads to an SHG enhancement in the energetically less favorable upright molecular arrangements, revealing a fundamental trade-off between interfacial thermodynamic stability and optical tensor activation. On non-centrosymmetric substrates, we uncover a characteristic, Fano-like tilt-sign asymmetry driven by constructive and destructive interference between the molecular and substrate amplitudes. These results establish actionable design rules for the predictive high-throughput screening of molecular adsorbates as SHG-active interfacial defects.

%% ============================================================
\section{Results}

%% ============================================================

%In the following, we present the modular framework, reconstructing the nonlinear susceptibility of van-der-Waals-bonded organic/inorganic interfaces from the intrinsic properties of the isolated constituents and their mutual arrangement (Sec.~\ref{sec:master}). We then detail the microscopic evaluation of the molecular hyperpolarizability tensor via a sum-over-states formulation and its two-level limit (Sec.~\ref{sec:chimol}), followed by the explicit treatment of interfacial electric-field-induced second-harmonic (EFISH) contributions arising from permanent dipole moments and substrate screening (Sec.~\ref{sec:efish}). Finally, we summarize the computational settings adopted to parametrize our model (Sec.~\ref{sec:protocol}).

\subsection{Effective susceptibility and symmetry activation}
\label{sec:master}

The effective second-order susceptibility of the hybrid interface is decomposed into three distinct contributions:
\begin{equation}\label{eq:chi2eff}
  \chihyb = \chiML + \chimol + \chiEF,
\end{equation}
where $\chiML$ is the intrinsic second-order response of the substrate dictated by its point-group symmetry (Table~\ref{tab:substrate_symmetry}), and $\chimol$ is the molecular hyperpolarisability tensor rotated from the molecular frame $(a,b,c)$ to the laboratory frame $(i,j,k)$:
\begin{equation}\label{eq:chimol_rot}
\chi^{(2)}_{\mathrm{mol},ijk}(\theta,\phi,\psi) = N_s R_{ia}R_{jb}R_{kc} \beta^{\mathrm{mol}}_{abc}.
\end{equation}
Here, $\beta^{\mathrm{mol}}_{abc}$ represents the microscopic first hyperpolarizability tensor evaluated in the unperturbed molecular coordinate system, $N_s$ is the surface number density of adsorbed molecules, $R(\theta,\phi,\psi)$ is the standard Euler rotation matrix parameterized by the molecular tilt ($\theta$), azimuthal ($\phi$), and roll ($\psi$) angles, and summation over repeated Cartesian indices is implied.
The third term in Eq.~\eqref{eq:chi2eff}, 
\begin{equation}
    \chiEF = \chi^{(3)}_{sub,ijkl}F^{\mathrm{dip}}_l
\end{equation}
is the electric‑field‑induced second‑harmonic generation (EFISH) contribution arising from the static field generated by the molecular permanent dipole moments:
\begin{equation}\label{eq:efish_field}
F^{\mathrm{dip}}_l = \frac{N_s \left[ R(\theta,\phi,\psi)\boldsymbol{\mu}_0 \right]_l}{\varepsilon_l},
\end{equation}
where $\boldsymbol{\mu}_0$ is the ground-state static dipole vector of the isolated molecule, and $\varepsilon_l$ represents the directional components of the dielectric screening tensor of the substrate ($\varepsilon_z=\varepsilon_\perp$, $\varepsilon_{x,y}=\varepsilon_\parallel$). 

For given incident and detected polarization directions, $\hat{\mathbf{e}}_\alpha$ and $\hat{\mathbf{e}}_\beta$, respectively, the SHG intensity scales as $I_{2\omega}^{(\alpha\to\beta)} \propto
|A_{2\omega}^{(\alpha\to\beta)}|^2$, where the total coherent amplitude is the sum of three contributions:

\begin{equation}\label{eq:amplitude}
\begin{aligned}
A_{2\omega}^{(\alpha\to\beta)}
&=
\hat{\mathbf{e}}_\beta \cdot \chihyb :
\hat{\mathbf{e}}_\alpha\hat{\mathbf{e}}_\alpha
\\
&=
A_{\mathrm{sub}}^{(\alpha\to\beta)}
+
A_{\mathrm{mol}}^{(\alpha\to\beta)}
+
A_{\mathrm{EFISH}}^{(\alpha\to\beta)} .
\end{aligned}
\end{equation}

with 
\begin{align}
A_{\mathrm{sub}}^{(\alpha\to\beta)}
&=
\hat{\mathbf{e}}_\beta \cdot \chiML :
\hat{\mathbf{e}}_\alpha\hat{\mathbf{e}}_\alpha,
\\
A_{\mathrm{mol}}^{(\alpha\to\beta)}
&=
\hat{\mathbf{e}}_\beta \cdot \chimol :
\hat{\mathbf{e}}_\alpha\hat{\mathbf{e}}_\alpha,
\\
A_{\mathrm{EFISH}}^{(\alpha\to\beta)}
&=
\hat{\mathbf{e}}_\beta \cdot \chiEF :
\hat{\mathbf{e}}_\alpha\hat{\mathbf{e}}_\alpha.
\end{align}
This coherent summation allows for direct interference between the molecular and substrate channels as discussed in Sec.~\ref{sec:results_interference}.

\subsection{Molecular hyperpolarisability: sum-over-states expression and two-level approximation}
\label{sec:chimol}
 
The full molecular hyperpolarisability tensor is obtained from the Orr--Ward sum-over-states (SOS) expansion~\cite{Orr1971}:

\begin{equation}\label{eq:sos_main}
  \beta_{ijk}(-2\omega;\omega,\omega) =
  \frac{1}{\hbar^2}
  \sum_{n,m\neq g}
  \frac{\mu^{i}_{gn}\;\bar{\mu}^{j}_{nm}\;\mu^{k}_{mg}}
       {(\omega_{ng}-2\omega-i\Gamma)(\omega_{mg}-\omega-i\Gamma)},
\end{equation}

where $\mu^{i}_{gn}=\langle g|\hat{\mu}_i|e_n\rangle$ represents the
ground-to-excited-state transition dipole moment, $\bar{\mu}^{j}_{nm}=\langle e_n|\hat{\mu}_j|e_m\rangle - \delta_{nm}\langle g|\hat{\mu}_j|g\rangle$ are the dipole-subtracted excited-to-excited matrix elements, $\omega_{ng}$ are the excitation frequencies, and $\Gamma$ is a phenomenological linewidth. The subtraction of the ground-state expectation value enforces $\beta\equiv0$ for centrosymmetric configurations, as required by Neumann's principle. In this work, the SOS in Eq.~\eqref{eq:sos_main} serves as a quasi-static, off-resonant descriptor to evaluate tensor symmetries and relative magnitudes across various adsorbate configurations. 
 
To guide physical intuition, when a single bright intramolecular transition dominates the optical response, as in push-pull chromophores~\cite{mili+03jpcb,nade+24rscad,thei+26apr},  Eq.~\eqref{eq:sos_main} simplifies to the two-level (2LVL) approximation:

\begin{equation}\label{eq:twolevel}
  \beta^{\mathrm{2LVL}} \approx
  \frac{|\mu_{ge}|^2\,\Delta\mu_{eg}}
       {(\tilde{E}_{eg}-2\hbar\omega-i\Gamma)
        (\tilde{E}_{eg}-\hbar\omega-i\Gamma)},
\end{equation}

where $\tilde{E}_{eg}$ is the substrate-renormalized excitation energy, $\mu_{ge}$ is the transition dipole moment, and $\Delta\mu_{eg}=\langle e|\hat{\mu}|e\rangle -
\langle g|\hat{\mu}|g\rangle$ is the net difference between the permanent dipole moments in the molecule in its excited- and ground-state. Eq.~\eqref{eq:twolevel} highlights a fundamental quantum-chemical trade-off: maximizing $\Delta\mu_{eg}$ requires significant intramolecular charge redistribution, which inherently reduces the spatial overlap of the frontier orbitals and thus reduces $|\mu_{ge}|^2\Delta\mu_{eg}$. While the full SOS expression (Eq.~\ref{eq:sos_main}) is utilized for all calculations, this two-level baseline serves as a conceptual guide for interpreting the results of our screening.

\subsection{The EFISH Coupling Mechanism}\label{sec:efish}

The EFISH contribution $\chiEF$ fundamentally alters the interfacial response by exploiting the third-order susceptibility of the substrate ($\chithree$). Since fourth-rank tensors are symmetry-allowed in all crystal structures, including centrosymmetric environments where intrinsic second-harmonic generation ($\boldsymbol{\chi}^{(2)}_{\mathrm{ML}}$) vanishes, an interfacial static electric field $\mathbf{F}^{\mathrm{static}}$ breaks local inversion symmetry, activating an effective second-order response: $\chi^{(2)}_{\mathrm{EFISH}} \propto \chi^{(3)} \mathbf{F}^{\mathrm{static}}$. The physical mapping of this coupling is highly sensitive to the adsorption geometry. As shown in Eq.~\eqref{eq:efish_field}, the static field components are determined by the projection of the rotated molecular dipole vector $\boldsymbol{\mu}_0$ onto the in-plane and out-of-plane components of the substrate. For a molecule with an in-plane permanent dipole moment along its long axis, changing the configuration from flat-lying ($\theta=0^\circ$) to upright ($|\theta|=90^\circ$) continuously transfers the electrostatic bias from an in-plane field ($F_{x,y}^{\mathrm{dip}} \propto \cos\theta$) to an out-of-plane field ($F_z^{\mathrm{dip}} \propto \sin\theta$). Consequently, varying the molecular tilt angle simultaneously rotates the intrinsic hyperpolarizability tensor $\chimol$ and systematically redistributes the EFISH weight across different polarization channels, directly coupling the geometric and chemical degrees of freedom.

Predicting $\chiEF$ is conceptually straightforward, as it depends on the ground-state molecular dipole moment, the interfacial dielectric environment, and the $\chithree$ tensor of the substrate. Since experimental values for $\chithree$ components are rarely available for isolated monolayers, we treat the relative magnitude of the third-order background as a phenomenological scaling descriptor; the adopted substrate tensor conventions and sensitivity analysis are reported in Section~S2 and Fig.~S1 of the Supporting Information. In practice, this channel is most relevant in centrosymmetric substrates, where it provides the primary mechanism for nonlinear optical activation. On non-centrosymmetric substrates, EFISH acts as a minor electrostatic perturbation superimposed on the dominant $\chiML$ response.

\begin{figure}%[htbp]
  \centering
  \includegraphics[width=0.5\linewidth]{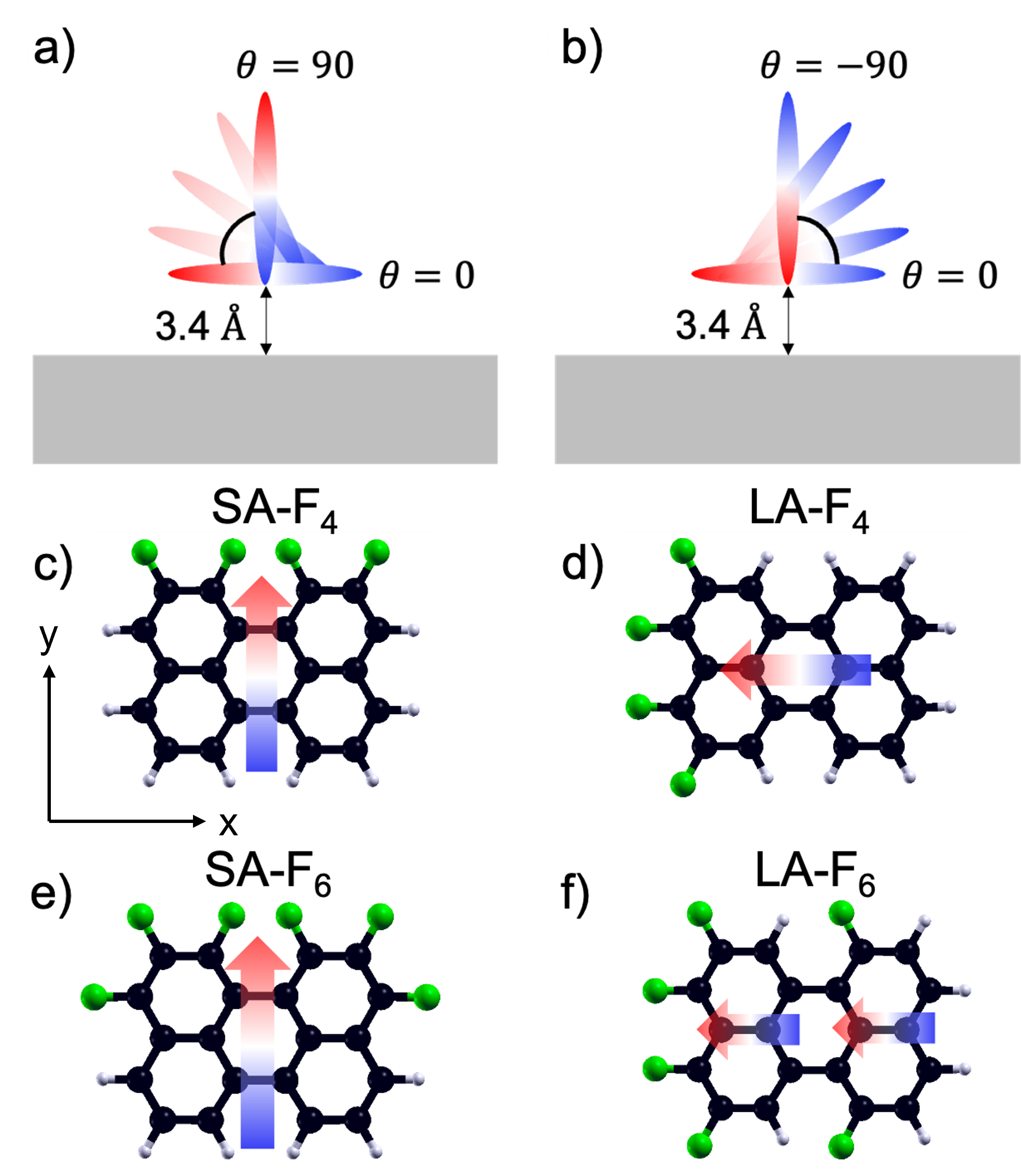}
  \caption{Schematic overview of adsorbate--substrate geometries for (a) positive and (b) negative tilt angles of the molecule, assuming a fixed interlayer distance of 3.4 \AA{}. (c--f) Fluorinated perylene derivatives indicating the orientation of the resulting ground-state permanent dipole moment (colored arrows) relative to the molecular axes: (c)~$\text{SA-F}_4$ and (d)~$\text{LA-F}_4$ isomers, with their dipole along the short axis (SA) and long axis (LA), corresponding to the zigzag and armchair directions oriented along $y$ and $x$, respectively; (e)~$\text{SA-F}_6$ and (f)~$\text{LA-F}_6$ isomers. Carbon, hydrogen, and fluorine atoms are depicted in black, white, and green, respectively.}
  \label{fig:tilt_geometry-systems}
\end{figure}

%% ============================================================
%\section{Results}
%\label{sec:results}
%% ============================================================

\subsection{Polar Molecular Adsorbates and Substrate Classes}
\label{sec:results_setup}

We consider a set of idealized perylene derivatives with a static dipole moment driven by asymmetric partial fluorination. Specifically, we investigate two structural isomers bearing a tetrafluorinated (F$_4$) substitution pattern: $\text{SA-F}_4$, with the dipole oriented along the short molecular axis (SA, zigzag direction, Fig.~1c), and $\text{LA-F}_4$, with the dipole oriented along the long molecular axis (LA, armchair direction, Fig.~1d). This functionalization generates a permanent dipole moment along the $x$- and $y$-direction, respectively. The spatial tilt scan for the LA-isomer (SA-isomer) corresponds to an out-of-plane rotation $R_y(\theta)$ ($R_x(\theta)$) about the $y$-axis ($x$-axis). When physisorbed flat on a substrate, both configurations possess C$_s$ point-group symmetry, yielding a non-zero hyperpolarizability tensor $\beta^{\mathrm{mol}}$ from our SOS reconstruction.

To assess interfacial coupling, each isomer is screened against the three representative 2D substrate classes categorized by their point-group symmetry (Table~\ref{tab:substrate_symmetry}). Focusing on MoS$_2$ as a prototypical member of the transition-metal dichalcogenide family, we associate the D$_{3h}$ group with the non-centrosymmetric monolayer phase (2H-MoS$_2$), D$_{3d}$ with centrosymmetric bilayers (e.g., AA'-stacked MoS$_2$~\cite{patr-cocc26arxiv}), and C$_{3v}$ with intrinsically polar Janus sheets (e.g., MoSSe)~\cite{bao+26prb}. For consistency across different substrate symmetries, we retain the dielectric screening constants specified in Sec.~\ref{sec:protocol},  which capture the essential polarization environment without altering the overarching qualitative symmetry trends or phase-matching behavior. As summarized in Table~\ref{tab:substrate_symmetry} and detailed in Ref.~\cite{patr-cocc26arxiv}, non-centrosymmetric monolayers are characterized by several non-zero elements of $\chi^{(2)}$. In contrast, centrosymmetric $D_{3d}$ bilayers satisfy $\chiML=0$, while retaining non-zero third-order susceptibilities ($\chithree \neq 0$), providing a clear, background-free platform to isolate molecule-induced EFISH activation from the intrinsic nonlinear response of the substrate.

\begin{table*}
  \centering
  \small
  \caption{Non-zero independent $\chi^{(2)}_{\mathrm{sub}}$ and $\chi^{(3)}_{\mathrm{sub}}$ tensor components for the investigated substrate symmetry classes (assuming standard Cartesian alignment with $y$ along a mirror plane for $D_{3h}/C_{3v}$). Permutations generated by index permutation symmetries are omitted for simplicity.}
  \label{tab:substrate_symmetry}
  \begin{tabular}{llll}
    \toprule
    Substrate Type & Symmetry Group & $\chi^{(2)}_{\mathrm{sub}} \neq 0$ Components & $\chi^{(3)}_{\mathrm{sub}} \neq 0$ Components \\
    \midrule
    Bilayer & $D_{3d}$ & none & $\chi^{(3)}_{xxxx}$, $\chi^{(3)}_{xxyy}$, $\chi^{(3)}_{xxzz}$, $\chi^{(3)}_{zzxx}$, $\chi^{(3)}_{zzzz}$ \\
    \addlinespace
    Monolayer & $D_{3h}$ & $\chi^{(2)}_{yyy}$, $\chi^{(2)}_{yxx}$, $\chi^{(2)}_{xxy}$, $\chi^{(2)}_{xyx}$ & $\chi^{(3)}_{xxxx}$, $\chi^{(3)}_{xxyy}$, $\chi^{(3)}_{xxzz}$, $\chi^{(3)}_{zzxx}$, $\chi^{(3)}_{zzzz}$ \\
    \addlinespace
    Janus Monolayer & $C_{3v}$ & Same as $D_{3h} + \chi^{(2)}_{zzz}$, $\chi^{(2)}_{zxx}$, $\chi^{(2)}_{xxz}$ & Same as $D_{3h}$ \\
    \bottomrule
  \end{tabular}
\end{table*}

\subsection{Adsorption on a Centrosymmetric Substrate}
\label{sec:results_bilayer}

\begin{figure*}
  \centering
    \includegraphics[width=1.0\linewidth]{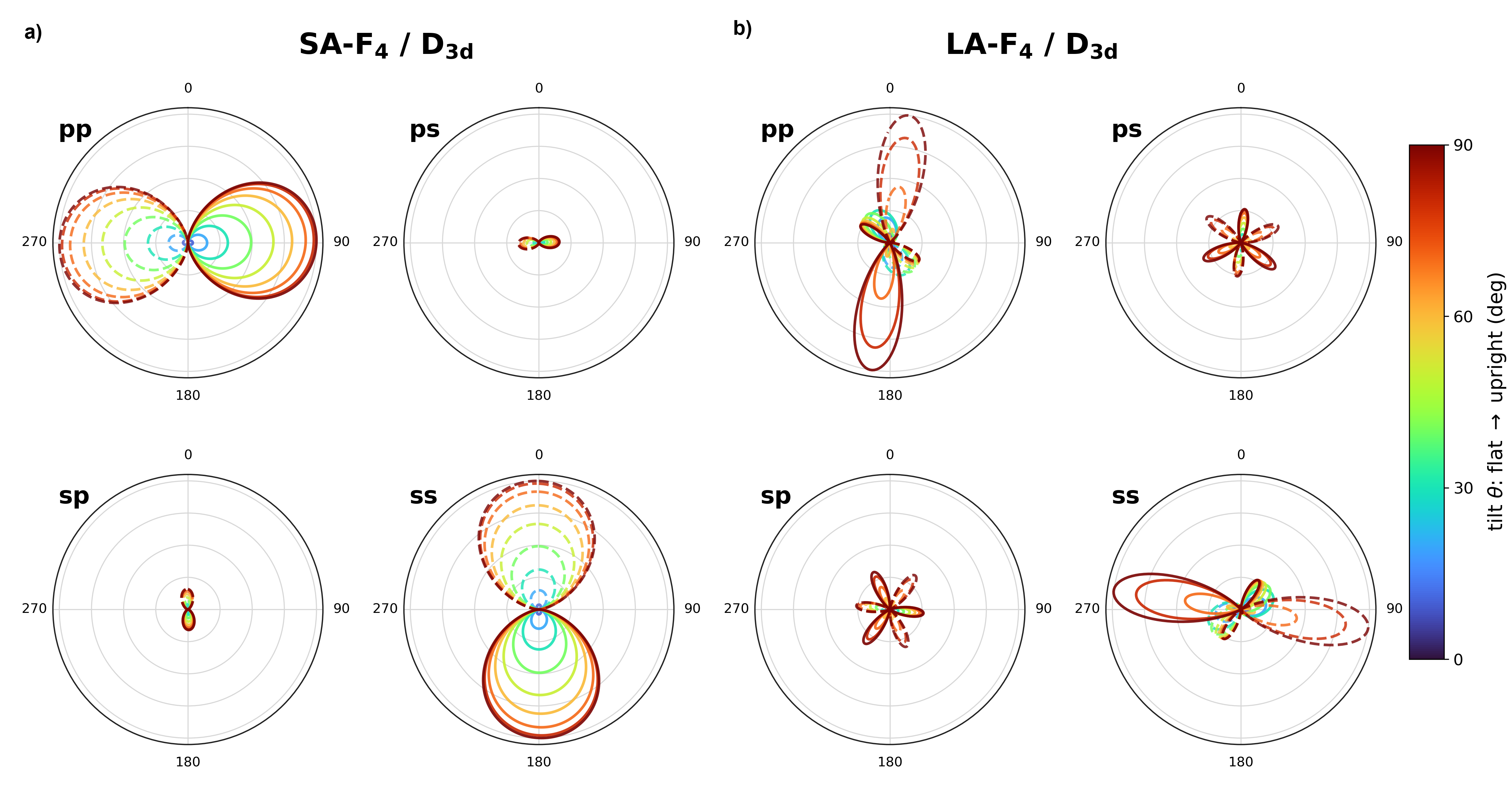}
  \caption{In-plane SHG polar plots generated by $pp$-, $ps$-, $sp$-, and $ss$-polarized light impinging a) the SA and b) the LA F$_4$-perylene isomer adsorbed on a centrosymmetric D$_{3d}$ substrate. The angular coordinate denotes the polarization angle $\phi \in \{0^{\circ} -360^{\circ}\}$ of the optical field, while the radial distance from the origin corresponds to the SHG intensity. The color scale tracks the molecular tilt angle $\theta$, stepping from a completely flat geometry ($\theta = 0^{\circ}$, dark blue) to a vertical standing conformation ($\theta = 90^{\circ}$, red). Solid and dashed curves serve as phase indicators, denoting positive and negative real amplitudes, respectively.}
  \label{fig:bilayer_inplane}
\end{figure*}

To isolate the molecular pathways and interfacial projection rules without background substrate interference, we first examine physisorption on a centrosymmetric bilayer ($D_{3d}$ group). As shown in Table~\ref{tab:substrate_symmetry}, the intrinsic second-harmonic response of the host matrix vanishes ($\chiML=0$), implying that net contributions to $\chi^{(2)}$ come exclusively from the molecular channels $\chimol$ or $\chiEF$. Under these background-free conditions, the LA ans SA isomers exhibit profoundly different nonlinear optical fingerprints (Fig.~\ref{fig:bilayer_inplane}). 

Under in-plane polarization configurations, denoted here as $ij$, where $i,j \in \{p,s\}$ specify the output second-harmonic ($2\omega$) and input fundamental ($\omega$) polarization states, respectively ($pp$, $ps$, $sp$, $ss$), the nonlinear response of the SA isomer exhibits a stable, monotonic evolution throughout the entire tilt sweep. (Fig.~\ref{fig:bilayer_inplane}a). In the dominant $pp$ and $ss$ channels, the SHG signal maps symmetric, two-lobed patterns aligned along the principal axes. As the molecule rotates out-of-plane, assuming upright orientation ($\theta \to 90^\circ$), these lobes expand uniformly without changes in shape or angular orientation, maximizing at an intensity baseline of 0.252. Across this trajectory, SHG emission preserves a uniform, positive real amplitude (solid lines in Fig.~\ref{fig:bilayer_inplane}a). This behavior demonstrates a smooth geometric mapping where the progressive out-of-plane projection of the primary molecular hyperpolarizability dictates the overall signal scaling.

The LA isomer exhibits a far more complex evolution (Fig.~\ref{fig:bilayer_inplane}b). Rather than scaling monotonically, its intensity profiles undergo drastic modifications during the tilt sweep, with the absolute maximum intensity at $\theta = 90^\circ$ being nearly four times weaker than the SA counterpart. This finding reveals that an upright geometry does not universally optimize the nonlinear optical output.  Crucially, azimuthal polar plots for the LA isomer reveal clear boundaries marked by sign reversals of the real SHG amplitude (solid \textit{vs.} dashed lines). These quadrant-dependent phase inversions demonstrate that tilting the LA molecule forces the driven electronic polarization to oscillate out-of-phase relative to the laboratory frame, inducing a discrete $180^\circ$ phase shift in the emitted second-harmonic wave. Identifying these phase boundaries is vital for molecular design rules, as negative real amplitudes give rise to destructive interference when coupled to a non-centrosymmetric substrate.

To clarify the microscopic origin of these contrasting profiles, we map the SHG patterns to the orientation-dependent tensor projections of $\beta^{\text{mol}}$. The clean, two-lobed topology and monotonic scaling of the SA configuration (Fig.~\ref{fig:bilayer_inplane}a) stem from its symmetric fluorination pattern, which concentrates the dominant hyperpolarizability along a single axial component ($\beta^{\text{mol}}_{yyy}$). As the SA molecule tilts about the $x$-axis, this primary component is smoothly projected into the out-of-plane laboratory channels without mixing with orthogonal tensor elements. Furthermore, the induced charge distribution remains symmetric across the surface plane, enforcing a strict equivalence among the mixed cross-tensor projections ($\chi_{zxx} \approx \chi_{zyy}$). 

Conversely, the asymmetric fluorination of the LA isomer introduces pronounced electronic anisotropy, yielding non-negligible off-diagonal hyperpolarizability elements ($\beta_{xxx}$, $\beta_{xyy}$, and $\beta_{yxx}$). When the LA molecule undergoes spatial rotation about the $y$-axis, these off-diagonal tensor components project onto the laboratory interface at competing rates,  giving rise to the multi-lobed patterns and rapid phase oscillations [Fig.~\ref{fig:bilayer_inplane}(b)]. At specific azimuthal angles $\phi$, these distinct components undergo mutual cross-cancellation, explaining why the LA isomer exhibits a non-monotonic channel evolution and significant SHG intensity suppression compared to the SA configuration. These findings highlight that optimizing nonlinear optical activation depends less on maximizing the total dipole vector, and more on preventing destructive self-interference among competing molecular tensor projections.

\begin{figure*}[!t]
  \centering
    \includegraphics[width=1.0\linewidth]{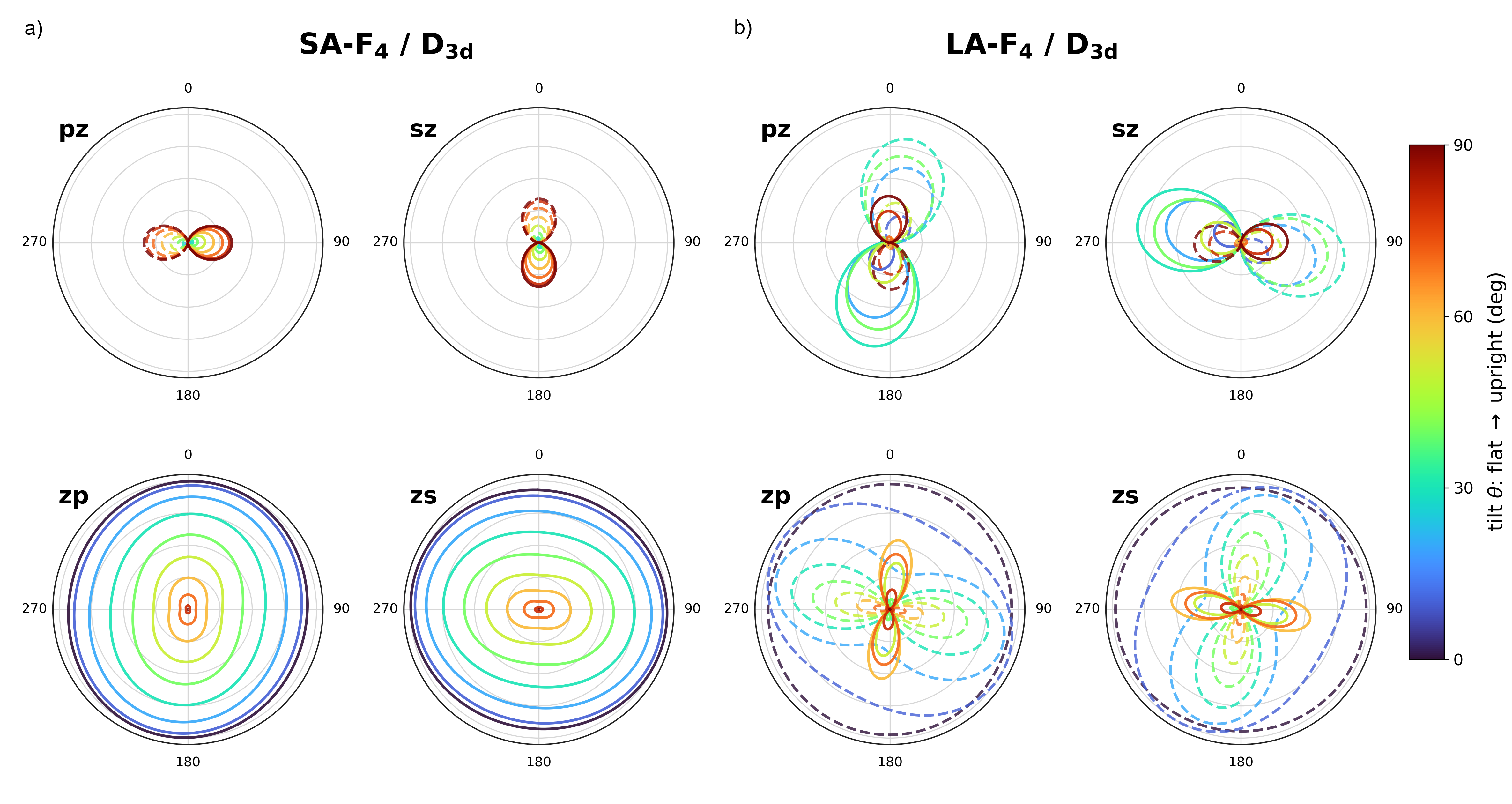}
  \caption{Mixed-$z$ polarization channels ($pz$, $sz$, $zp$, $zs$) on the centrosymmetric D$_{3d}$ bilayer substrate for (a) the SA-F$_4$ isomer (tilt $R_x(\theta)$ about the $x$-axis) and (b) LA-F$_4$ isomer (tilt $R_y(\theta)$ about the $y$-axis). The SA $zp/zs$ patterns maintain near-circular, isotropic profiles ($\chi_{zxx} \approx \chi_{zyy}$), whereas the LA channels exhibit multi-lobed structures and phase reversals (solid \textit{vs.} dashed lines) driven by strong hyperpolarizability anisotropy.}
  \label{fig:bilayer_mixedz}
\end{figure*}

To determine how the molecular point-group symmetries project onto the laboratory framework, we extend this phase-resolved analysis to the mixed-$z$ polarization pathways ($pz, sz, zp, zs$). Probing these channels highlights the geometric constraints governing out-of-plane nonlinear transport at the interface. As shown in Fig.~\ref{fig:bilayer_mixedz}(a) for the SA isomer, the $zp$ and $zs$ channels yield concentric, isotropic patterns across the entire spatial trajectory. This azimuthal invariance confirms the spatial equivalence of the out-of-plane cross-tensor components ($\chi_{zxx} \approx \chi_{zyy}$). As the SA molecule rotates out of a flat orientation ($\theta = 0^\circ$, dark blue) toward an upright configuration ($\theta = 90^\circ$, red), these rings contract uniformly toward the origin while maintaining isotropic symmetry.

In contrast, the LA isomer breaks this in-plane isotropy entirely [Fig.~\ref{fig:bilayer_mixedz}(b)]. Its $zp$ and $zs$ channels display heavily distorted oval shapes that evolve dynamically with the tilt angle $\theta$. Furthermore, these projections exhibit localized phase boundaries (solid \textit{vs.} dashed curves within the same azimuthal sweep), demonstrating that the off-diagonal hyperpolarizability elements ($\beta_{xxx}, \beta_{xyy}, \beta_{yxx}$) undergo asynchronous phase shifts when projected onto the surface normal.

This spatial mapping additionally reveals a fundamental physical asymmetry between channels where the out-of-plane component acts as the emitted second-harmonic field ($zp, zs$) and those where it acts as an incoming driving field ($pz, sz$). For the SA isomer, the $pz$ and $sz$ configurations generate highly directional, two-lobed patterns [Fig.~\ref{fig:bilayer_mixedz}(b)], qualitatively similar to those obtained for the in-plane $pp$ and $ss$ directions, respectively [ Fig.~\ref{fig:bilayer_inplane}(a)]. This behavior reflects a tensor projection mechanism: in $zp/zs$ configurations ($\chi^{(2)}_{z\lambda\mu}$), in-plane fundamental fields ($\lambda, \mu \in \{x,y\}$) drive a symmetric charge displacement radiating uniformly along the surface normal. Conversely, in the $pz/sz$ configurations ($\chi^{(2)}_{\lambda z\mu}$), the out-of-plane fundamental field directly drives the electron density along $z$, which must then radiate back through the anisotropic in-plane axes ($\lambda$) of the molecular backbone, restricting efficient radiation to specific azimuthal angles.

\begin{figure*}%[htbp]
  \centering
  \includegraphics[width=\linewidth]{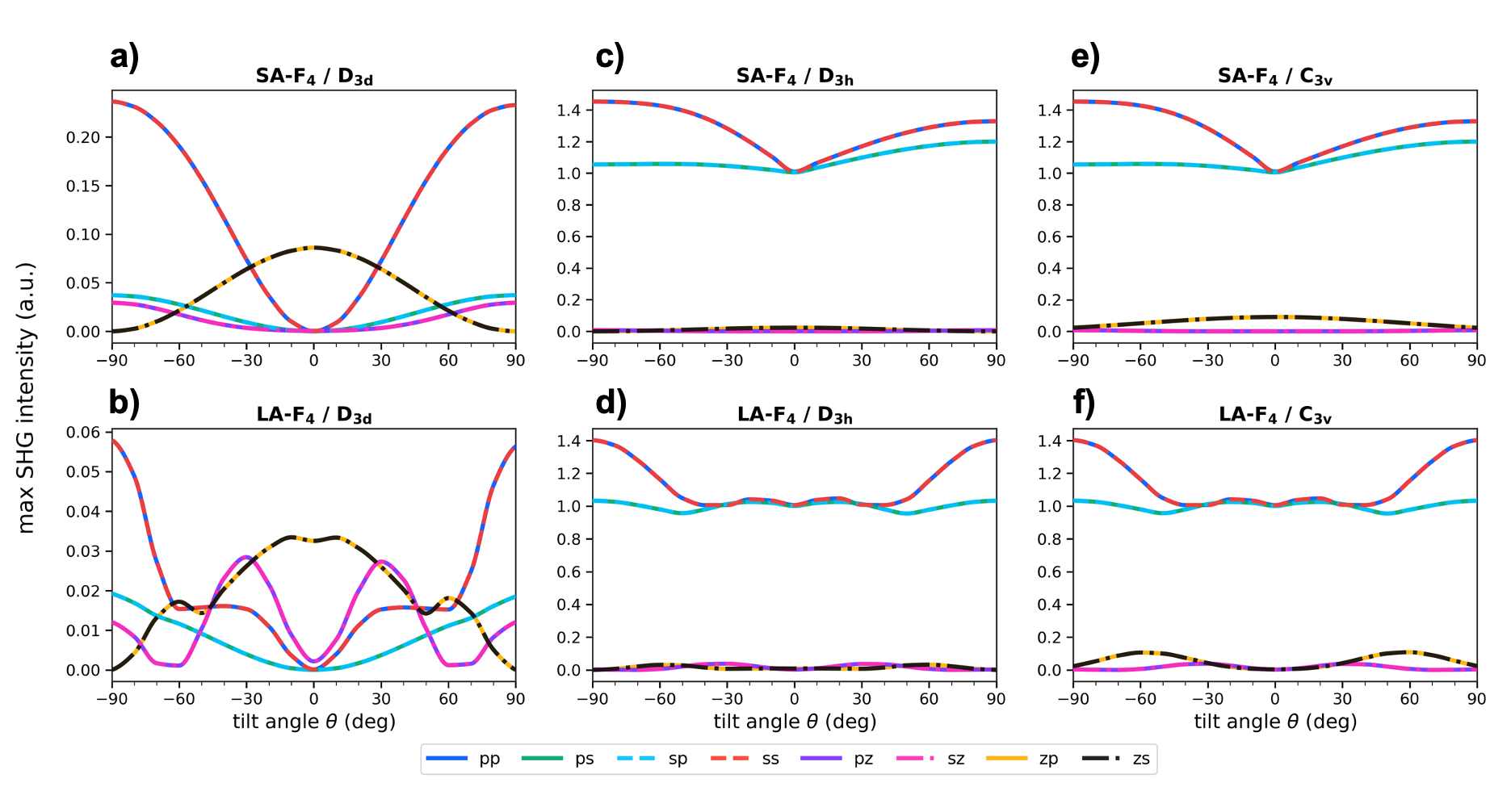}
  \caption{Channel-resolved maximum SHG intensity as a function of signed tilt angle $\theta\in[-90^\circ,90^\circ]$ for a) SA-F$_4$- and b) LA-F$_4$-perylene on a centrosymmetric substrate (D$_{3d}$ group), for c) SA-F$_4$- and d) LA-F$_4$-perylene on a non-centrosymmetric substrate (D$_{3h}$ group), and e) SA-F$_4$- and f) LA-F$_4$-perylene on a
C$_{3v}$ Janus-type substrate.}
  \label{fig:f4_channel_vs_tilt}
\end{figure*}

The macroscopic consequences of these distinct tensor mappings are reflected in the angle-dependent maximum SHG intensity profiles calculated for the considered adsorbate/substrate combinations (Fig.~\ref{fig:f4_channel_vs_tilt}). For the SA isomer, the tilt evolution reveals a parabolic decoupling between polarization channels [Fig.~\ref{fig:f4_channel_vs_tilt}(a)]. The in-plane $pp$ and $ss$ signals exhibit monotonic quadratic growth, peaking at the vertical limits ($\theta = \pm90^\circ$). Concurrently, the out-of-plane $zp$ and $zs$ channels scale inversely, reaching their maximum at a completely flat adsorption geometry ($\theta = 0^\circ$) before decaying smoothly to zero at upright orientations. 

Conversely, the LA isomer reveals a competitive multi-channel landscape marked by prominent, non-monotonic oscillations [Fig.~\ref{fig:f4_channel_vs_tilt}(b)]. The dominant in-plane $pp$ and $ss$ signals experience sharp drops at intermediate orientations, hitting distinct local minima near $\theta = \pm45^\circ$, where the mixed-$z$ channels ($zp$, $zs$, and $sz$) undergo substantial enhancement to become the locally dominant modes of nonlinear emission. This crossover can be understood as follows: While the in-plane driven $zp$ and $zs$ channels collapse into multi-lobed structures due to minor off-diagonal tensor mixing, the $pz$ and $sz$ configurations maintain remarkably stable two-lobed footprints. This stability occurs because rotating the LA framework about its short molecular axis ($y$) lifts its primary charge-transfer axis (embedding the dominant hyperpolarizability element $\beta_{xxx}$) directly into the $z$-direction. By coupling to a $z$-polarized fundamental photon, the $pz/sz$ channels exploit this single pathway cleanly,  avoiding the destructive cross-cancellations that plague the in-plane driven modes.

\subsection{Adsorption on Non-Centrosymmetric Substrates}
\label{sec:results_interference}

Assuming a non-centrosymmetric substrate, the nonlinear optical response of the hybrid interface changes substantially. On both the $D_{3h}$-symmetric monolayer and the $C_{3v}$-symmetric Janus sheet, several tensor elements of the second-order susceptibility are instrinsically non-zero (Table~\ref{tab:substrate_symmetry})~\cite{patr-cocc26arxiv,bao+26prb}. Consequently, the detected SHG intensity $I_{2\omega}$ can no longer be treated as isolated molecular emission, but requires a coherent, phase-dependent superposition of the substrate and molecular radiation fields:
\begin{equation}\label{eq:interference}
  I_{2\omega}(\theta) = |A_{\mathrm{sub}} + A_{\mathrm{mol}}(\theta)|^2 = |A_{\mathrm{sub}}|^2 + |A_{\mathrm{mol}}(\theta)|^2 + 2\,\mathrm{Re}\left[A_{\mathrm{sub}}^* A_{\mathrm{mol}}(\theta)\right],
\end{equation}
where $A_{\mathrm{sub}}=\hat{\mathbf{e}}_\beta\cdot\chiML:
\hat{\mathbf{e}}_\alpha\hat{\mathbf{e}}_\alpha$ and
$A_{\mathrm{mol}}(\theta)=\hat{\mathbf{e}}_\beta\cdot[\chimol(\theta)+\chiEF(\theta)]:
\hat{\mathbf{e}}_\alpha\hat{\mathbf{e}}_\alpha$ are the substrate and molecule-induced amplitudes emitted by the substrate and the molecular layer, respectively, for a given polarization channel.

The cross-term on the r.h.s. of Eq.~\eqref{eq:interference} acts as a phase-sensitive modulator. Because the molecular projection changes sign under opposite orientation angles [$A_{\mathrm{mol}}(-\theta) = -A_{\mathrm{mol}}(+\theta)$ for specific tensor components], this cross-term drives constructive interference on one tilt branch and destructive interference on the other, resulting in a characteristic $+\theta/-\theta$ intensity asymmetry in $I_{2\omega(\theta)}$. By analogy with the line shapes arising from the coupling of a discrete resonance to a continuum background, we term this behavior Fano-like interference, where the tilt angle $\theta$ serves as a physical detuning parameter in place of the photon energy.

As a consequence, on non-centrosymmetric substrates, the in-plane polarization channels ($pp$ and $ss$) never drop to zero~\cite{patr-cocc26arxiv}. Their baseline is pinned by the intrinsic susceptibility of the host monolayer ($|A_{\mathrm{sub}}|^2$), with the physisorbed molecular layer merely modulating a strong optical background. Comparing the two fluorinated isomers reveals that the activation of this Fano-like asymmetry depends heavily on the internal symmetry of the hyperpolarizability tensor.
On both $D_{3h}$ and $C_{3v}$ substrates, the SA isomer shows a pronounced $+\theta/-\theta$ magnitude asymmetry [Fig.~\ref{fig:f4_channel_vs_tilt}(c)-(f)]: its maximum SHG intensity climbs to 1.453 for $\theta = -90^{\circ}$, but only reaches 1.328 at $\theta = +90^{\circ}$. This divergence confirms that the phase-uniform tensor projection of the SA isomer locks into constructive or destructive alignment with the substrate background vector. On the other hand, the LA isomer maintains symmetric intensity profiles across both tilt branches [$\pm90^{\circ}$, Fig.~\ref{fig:f4_channel_vs_tilt}d,f)]. Since the spatially distributed, anisotropic hyperpolarizability of the LA molecule induces rapid, quadrant-dependent phase reversals [compare Fig.~\ref{fig:bilayer_inplane}(b)], the interference cross-term undergoes mutual self-cancellation when integrated over the full azimuthal sweep.

The $C_{3v}$ Janus substrate introduces additional non-zero out-of-plane tensor components ($\chi^{(2)}_{zzz}, \chi^{(2)}_{zxx}, \chi^{(2)}_{xxz}$)~\cite{bao+26prb}, elevating the mixed-$z$ baseline relative to the $D_{3h}$ host matrix. As summarized in Table~\ref{tab:results_summary}, the maximum SHG intensity at a completely flat adsorption geometry ($\theta = 0^\circ$) remains identical between the two substrates (1.006 for LA \textit{vs.} 1.010 for SA), revealing a regime of complete electrostatic and symmetry decoupling. Since a flat-lying molecule projects no intrinsic out-of-plane hyperpolarizability elements ($\boldsymbol{\beta}^{\mathrm{mol}}_{\perp} = 0$), it cannot couple with the vertical background fields of the substrate. Furthermore, because the local field-induced bias vector scales with the out-of-plane projection of the molecular frame ($F_z^{\mathrm{DC}} \propto \sin\theta \to 0$), the EFISH contribution vanishes at $\theta = 0^\circ$. Consequently, the low-tilt mixed-$z$ response remains entirely substrate-dominated, leaving the physisorbed layer completely blind to the broken out-of-plane symmetry of the Janus monolayer unless a structural tilt is introduced ($\theta > 0^\circ$).

\subsection{F$_6$-perylene Adsorbates}
\label{sec:f6}

To confirm that the failure of the ground-state permanent dipole moment as a predictive descriptor for SHG is not an artifact of specific F$_4$-substitution patterns, we extend our analysis to hexa-fluorinated perylene
(F$_6$-perylene) isomers. As shown in Fig.~\ref{fig:tilt_geometry-systems}(e)-(f), we apply an analogous functionalization scheme for LA and SA isomers. As a consequences of the expanded fluorination, the F$_6$-perylene series exhibits an identical spatial distribution of its primary dipole axes with an overall higher magnitude relative to F$_4$-functionalization: the SA isomer remains the more polar system ($|\mu|=5.77$\,D) compared to its LA counterpart ($|\mu|=4.96$\,D).

\begin{figure*}%[htbp]
  \centering
  \includegraphics[width=\linewidth]{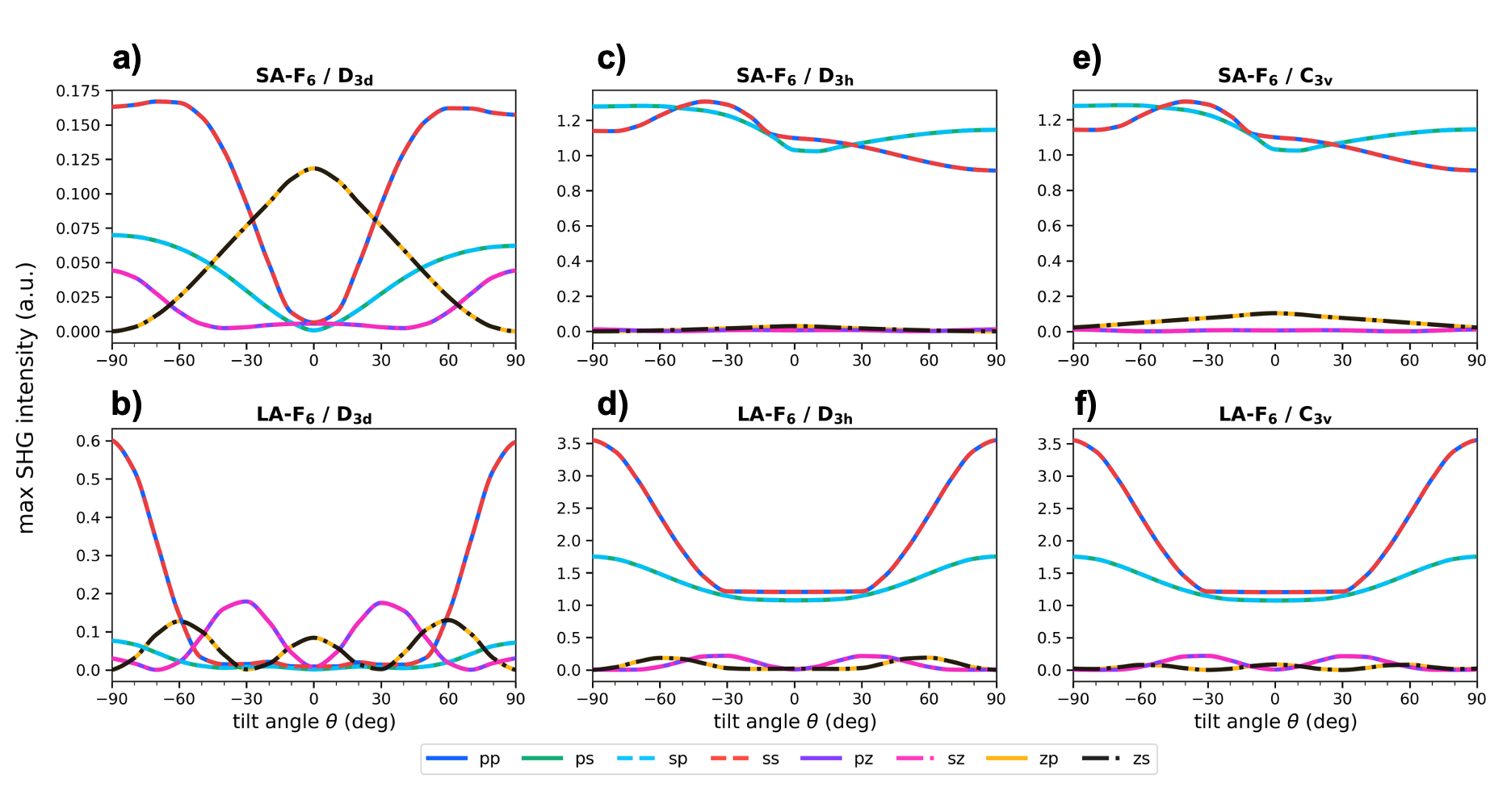}
  \caption{Maximum SHG intensity over all polarization channels as a function of signed tilt angle $\theta \in [-90^\circ, 90^\circ]$ for (a) SA-F$_6$ and (b) LA-F$_6$ isomers on the centrosymmetric D$_{3d}$ substrate, (c) SA-F$_6$ and (d) LA-F$_6$ on the non-centrosymmetric D$_{3h}$ monolayer, and (e) SA-F$_6$ and (f) LA-F$_6$ on the Janus-type substrate (C$_{3v}$ symmetry).}
  \label{fig:f6_channel_vs_tilt}
\end{figure*}

When evaluated on the background-free $D_{3d}$ bilayer substrate, the continuous signed-tilt sweeps in Fig.~\ref{fig:f6_channel_vs_tilt}(a)-(b) reveal a dramatic ranking reversal relative to electrostatic intuition. At higher vertical tilt angles ($\theta \to \pm 90^\circ$), the less polar $\mathrm{F}_6$-LA isomer outperforms its more polar SA counterpart in maximum SHG intensity by nearly a factor of four. Owing to the absence of intrinsic substrate $\chi^{(2)}$ components, the SHG intensity curves remain symmetric across the $\pm\theta$ branches, while the dominant emission channel are swapped from $ss$ at $+90^\circ$ to $pp$ at $-90^\circ$.

As summarized in Table~\ref{tab:results_summary}, this ranking reversal persists on non-centrosymmetric substrates belonging to $D_{3h}$ and $C_{3v}$ groups. Despite the intrinsic SHG background from the host matrix, the less polar $\text{F}_6$-LA isomer reaches a peak intensity of 3.554, while the highly polar $\text{F}_6$-SA variant remains between 1.144 and 1.278. Similar to its $\text{F}_4$ counterpart, the $\text{F}_6$-SA configuration displays a strong Fano-like tilt asymmetry due to coherent phase coupling with the substrate, whereas the LA profile maintains symmetric response across $\pm\theta$.

\begin{table*}%[!t]
  \centering
  \small
  \caption{Permanent dipole magnitude $|\boldsymbol{\mu}|$ and maximum SHG intensity (in model units, maximized over polarization channel and azimuth angle $\phi$) at flat ($\theta=0^\circ$) and upright ($\theta=\pm90^\circ$) orientations for the LA/SA F$_4$- and F$_6$-perylene isomers on the three model substrates with varying symmetry. For the $\mathrm{F}_4$ pair on $D_{3d}$, dipole magnitude correlates with SHG response, whereas the $\mathrm{F}_6$ pair exhibits a complete ranking reversal, demonstrating that ground-state polarity alone cannot predict nonlinear optical enhancement.}
  \label{tab:results_summary}
  \begin{tabular}{llcccc}
    \toprule
    Substrate symmetry & System & $|\mu|$ (D) & SHG ($\theta = 0^\circ$) & SHG ($\theta = +90^\circ$) & SHG ($\theta = -90^\circ$) \\
    \midrule
    D$_{3d}$ & LA-F$_4$ perylene & 3.01 & 0.033 & 0.056 & 0.058 \\
             & SA-F$_4$ perylene & 4.77 & 0.086 & 0.233 & 0.237 \\     \addlinespace
             & LA-F$_6$ perylene & 4.96 & 0.084 & 0.597 & 0.602 \\
             & SA-F$_6$ perylene & 5.77 & 0.118 & 0.157 & 0.163 \\
\hline
    D$_{3h}$ & LA-F$_4$ perylene & 3.01 & 1.006 & 1.402 & 1.402 \\
             & SA-F$_4$ perylene & 4.77 & 1.010 & 1.328 & 1.453 \\ \addlinespace
             & LA-F$_6$ perylene & 4.96 & 1.205 & 3.547 & 3.547 \\
             & SA-F$_6$ perylene & 5.77 & 1.100 & 1.146 & 1.279 \\
\hline
    C$_{3v}$ & LA-F$_4$ perylene & 3.01 & 1.006 & 1.402 & 1.402 \\
             & SA-F$_4$ perylene & 4.77 & 1.010 & 1.328 & 1.453 \\ \addlinespace
             & LA-F$_6$ perylene & 4.96 & 1.205 & 3.554 & 3.554 \\
             & SA-F$_6$ perylene & 5.77 & 1.100 & 1.144 & 1.278 \\
    \bottomrule
  \end{tabular}
\end{table*}

To ensure that this ranking reversal represents a physically robust phenomenon rather than an artifact of our model, we performed a numerical sensitivity check by globally scaling the non-zero substrate $\chi^{(2)}$ and $\chi^{(3)}$ tensor components (details in Figs.~S1-S18 and Table~S1). Decreasing the scale factor isolates the intrinsic molecular orientation dependence, while increasing it strengthens the substrate background and EFISH interference. Across all scaling regimes, the qualitative picture holds: while the $\text{F}_4$-isomers maintain a consistent hierarchy due to the strong anisotropy of the long-axis hyperpolarizability, the $\text{F}_6$-series undergoes a substrate-mediated ranking reversal driven by coherent phase interference. This confirms that the superior nonlinear optical performance of the less polar LA architecture is dictated by optimized hyperpolarizability tensor projections rather than the permanent dipole magnitude alone. Nevertheless, while this internal sensitivity check confirms the algebraic stability of our screening model, full \textit{ab initio} quantum mechanical calculations of the complete hybrid heterostructures remain the natural step forward to benchmark these predictions against interfacial hybridization~\cite{jaco+22acsanm}.

\section{Methods}
\label{sec:protocol}

\textbf{Structure generation and High-Throughput Screening:}
Substituted perylene derivatives are enumerated by identifying symmetry-inequivalent substitution patterns via canonical-SMILES~\cite{Han2025,Furness2025} deduplication. Initial geometries are generated with the \texttt{RDKit} cheminformatics toolkit ~\cite{RDKitBook2026} under a planarity-preserving embedding protocol. Centrosymmetric isomers are automatically filtered out and discarded from the screening pool.

\textbf{Electronic Structure Calculations:}
All \textit{ab initio} calculations were performed using the code \texttt{Octopus}~\cite{TancogneDejean2020}. Density-functional theory (DFT)~\cite{hohe-kohn64pr,kohn-sham65pr} calculations were carried out in a spin-unpolarized framework within the local-density approximation (LDA) for the exchange-correlation functional~\cite{perd-zung81prb} as implemented in HGH-LDA pseudopotentials~\cite{hart+98prb}, a real-space grid spacing of 0.24~\AA{}, and a spherical simulation box of radius 6.0~\AA{}. Singlet excited states were obtained from Casida's linear-response formulation~\cite{casi09jms} of time-dependent DFT (TDDFT)~\cite{rung-gros84prl}, using 100 empty states, a 30~eV transition-energy window, and including the first 100 excitations for SOS reconstruction. Further details are reported in the Supporting Information.

\textbf{Model Parameterization and Screening Descriptors:} Substrate screening was modeled using \texttt{LayerPCM}~\cite{Krumland2021b} as implemented in \texttt{Octopus}. Assuming monolayer MoS$_2$ with an effective thickness of 5.46~\AA{}, the in-plane and out-of-plane dielectric constants were set to $\varepsilon_{\parallel}=16.52$, $\varepsilon_{\perp}=10.08$, respectively~\cite{Krumland2021b}. These settings were applied in the seed TDDFT calculations to renormalize molecular excitation energies and to screen the permanent dipole in the EFISH descriptor. The tensor-level substrate response in the hybrid SHG model is treated separately as the normalized phenomenological $\chi^{(2)}_{\mathrm{sub}}/\chi^{(3)}_{\mathrm{sub}}$ model described in the Supporting Information.

%\textbf{Geometric Tilt Descriptor:}
The molecular tilt angle $\theta$ is defined over the range $[-90^\circ,90^\circ]$, where $\theta=0^\circ$ corresponds to the flat-lying physisorbed configuration at a fixed equilibrium distance of 3.4~\AA{} above the basal plane, and $|\theta|=90^\circ$ denotes the upright setting [Fig.~\ref{fig:tilt_geometry-systems}(a)-(b)]. To establish a clean descriptor-level diagnostic that decouples geometric frame transformations from electronic structural changes, we adopt a rigid-rotation protocol: the molecular hyperpolarizability tensor is computed for the reference $\theta=0^\circ$ geometry and analytically transformed into the laboratory frame using Eq.~\eqref{eq:chimol_rot}.

%\textbf{Substrate Normalization Baseline:}
To isolate the physical interplay between interfacial interference and orientation trends from specific absolute experimental susceptibilities, all substrate tensor amplitudes are normalized relative to the leading in-plane non-centrosymmetric component $\chi^{(2)}_{yyy}=1$ for the $D_{3h}$ symmetry group. All computed SHG intensities and interference profiles are reported in these normalized units. Using monolayer MoS$_2$ as a representative benchmark substrate, this scaling preserves all qualitative symmetry rules, phase dependencies, and relative channel activations while providing a transferrable framework applicable across a broad family of non-centrosymmetric two-dimensional materials.

%% ============================================================
\section{Discussion}
\label{sec:discussion}
%% ============================================================

The comparative analysis of the nonlinear optical response of two fluorinated perylene isomer series on three substrate symmetry classes demonstrates that SHG at low-dimensional hybrid organic/inoranic interfaces is governed by a non-trivial interplay between local symmetry breaking via molecular adsorption and coherent, phase-resolved interference between the nonlinear optical responses of adlayer and substrate. Crucially, these findings dismantle the conventional reliance on static polar descriptors, proving that the ground-state permanent dipole magnitude is an unreliable proxy for determining macroscopic nonlinear optical activity. This is clarified by the $\text{F}_6$-perylene series, where the less polar $\text{F}_6$-LA isomer outshines its highly polar $\text{F}_6$-SA variant by almost a factor of four at vertical tilt limits across all examined substrates. The governing criterion to maximize SHG is encoded in the full orientation-resolved tensor contraction, with the molecular tilt coordinate simultaneously dictating the spatial rotation of $\beta^{\mathrm{mol}}$ into the laboratory framework and reshaping the electrostatic weight of the EFISH. Optimizing second-order optical behavior therefore demands a unified multi-variable framework where the full anisotropic molecular tensor topology and the spatial dipole orientation are concurrently assessed. 

Based on this finding, the optimal molecular configuration remains critically determined by the point-group symmetry of the substrate. As shown for the $\text{F}_4$-perylene series, while the SA isomer dominates on the background-free, centrosymmetric $D_{3d}$ substrate, the LA isomer exhibits a superior nonlinear response on non-centrosymmetric matrices. This divergence is mediated by the phase-sensitive cross-term $2\,\mathrm{Re}[A_{\mathrm{sub}}^* A_{\mathrm{mol}}(\theta)]$, which determines the constructive or destructive nature of molecular-substrate signal interference. 

To assess the feasibility of these design rules, we performed fully atomistic, first-principles calculations of 16 representative interfaces, considering the geometric limits of flat-lying ($\theta = 0$) versus upright molecular adsorption configurations ($\theta = 90^\circ$) and focusing on the formation energy and the ground-state charge transfer that could potentially challenge our ansatz. Our results, detailed in the Supporting Information (Section~S4), show that the flat-lying configurations adsorb substantially more favorably (by more than 1~eV) than the upright orientation across all substrates. Despite this energetic penalty, adsorption energies of vertically oriented adlayers remain well below thermal energy at room temperature, suggesting the experimental feasibility of these configurations under appropriate kinetic growth conditions. Importantly, the ground-state charge transfer remains negligible at all considered interfaces, with maxima of $+0.013~e$ in the horizontal configuration and $-0.024~e$ in the vertical orientation. These findings provide strong evidence of physisorption, thereby fully justifying the assumptions behind this model. In particular, these conditions ensure that the molecular hyperpolarizability tensor $\boldsymbol{\beta}^{\mathrm{mol}}$ remains substantially unperturbed by interfacial interactions without significant charge redistribution regardless of the specific orientation and substrate composition.

\section{Conclusions and Outlook}

In summary, this work establishes a substrate-aware framework for predicting and engineering SHG at low-dimensional hybrid interfaces. Moving beyond traditional screening protocols that rely on simple scalar proxies, we have demonstrated that the permanent ground-state dipole moment is a fundamentally unreliable descriptor for predicting macroscopic nonlinear optical activity. Instead, the net coherent response is governed by a multi-variable parameter space in which the phase-resolved projection of the full molecular hyperpolarizability tensor couples intimately with host substrate symmetry and a phase-sensitive, Fano-like interfacial interference cross-term. By establishing the exact linkages between functionalization topology, dynamic spatial orientation, and active substrate backgrounds, these insights elevate SHG from a passive diagnostic tool to a predictive design paradigm.

Based on these foundational physical principles, we formulate four central design rules for engineering SHG-active non-centrosymmetric hybrid interfaces:
\begin{itemize}
  \item \textbf{Rule 1 -- Optimize the Quantum-Mechanical $\chi^{(2)}$ Numerator.} 
Candidate molecules must be screened using their full, phase-resolved hyperpolarizability tensor components rather than ground-state dipoles or simplified single-axis charge-transfer models. In multi-substituted halogenation topologies, second-order responses are driven by multi-directional polarization pathways rather than a unidirectional donor--acceptor charge-transfer axis. Consequently, reliable predictions require contracting the full molecular tensor $\beta^{\mathrm{mol}}(\theta, \phi)$ into the target laboratory polarization channels at the anticipated interfacial tilt angle.
  \item \textbf{Rule 2 -- Tune the $\chi^{(2)}$ Energy Denominator via Band Alignments.}
Molecular electronic transitions should be explicitly modulated to approach optical resonance. Aligning a molecular mid-gap state near the laser driving frequency ($\omega$) or the second-harmonic emission ($2\omega$) triggers a resonant enhancement of several orders of magnitude. This boost is driven by the structural minimization of the state-energy denominators ($\hbar\omega_{eg} - \hbar\omega$ and $\hbar\omega_{eg} - 2\hbar\omega$) within the SOS perturbative expansion.
  \item \textbf{Rule 3 -- Group-Theory-Driven Substrate Selection.}
Point-group symmetry must guide the adsorbate-substrate matching.  For centrosymmetric layers ($\chiML=0$), the molecular adlayer represents the sole SHG source: engineering efforts should therefore maximize the parallel $pp$/$ss$ projections of $\chimol+\chiEF$ at large tilt angles. Conversely, for non-centrosymmetric hosts ($\chiML \neq 0$), screening protocols must prioritize molecular configurations that emit in phase with the substrate background amplitude ($A_{\mathrm{sub}}$) in the target channel to prevent signal self-extinction.
  \item \textbf{Rule 4 -- Exploit Molecular Orientation as a Dynamical Knob.}
The signed tilt ($\theta$) and azimuthal rotation ($\phi$) must be leveraged as independent structural switches. On non-centrosymmetric substrates, the sign of the tilt breaks intensity degeneracy, yielding a pronounced Fano-like asymmetry between $+\theta$ and $-\theta$ branches while enabling polarization-channel inversion ($pp\leftrightarrow ss$). Since the orientation-dependent trajectory of $\chimol$ and $\chiEF$ is intrinsically linked to the functionalization pattern, both must be optimized jointly.
\end{itemize}

The electrostatic and perturbative model developed here is uniquely tailored for the high-throughput screening of hybrid interfaces for maximized second-order nonlinear optical signals. Thanks to its modular architecture and physics-informed approach, this framework can be naturally extended to higher perturbative orders, unlocking deeper physical insights and predictions of complex low-dimensional materials. Ultimately, this engine lays the foundation for data-driven discovery and machine learning models optimized for material diagnostics and next-generation nonlinear optoelectronics. Concurrently, subtle effects arising from the weak molecule-substrate interactions outlined in our fully atomistic DFT analysis will be comprehensively disentangled and interpreted in dedicated upcoming studies, further validating and complementing the robust screening framework proposed herein.

\backmatter

\bmhead{Supplementary information}
The structural parameters, analytical derivations, and \textit{ab initio} validations of the proposed model are reported in the Supporting Information. Section~S1 provides the explicit analytical reconstruction of the molecular hyperpolarizability tensors. Section~S2 outlines the computational parameters for both the screening engine and the first-principles electronic structure calculations of isolated molecules. Section~S3 includes additional polarization-resolved SHG polar patterns alongside a comparative analysis using pristine perylene signatures. Section~S4 presents fully atomistic, first-principles calculations on 16 selected heterostructures, detailing ground-state adsorption energetics and interfacial charge transfer.

\bmhead{Funding}
 This work was funded by the German Research Foundation, Project Number 398816777, CRC 1375 ``NOA'', subproject A8.

\bmhead{Data Availability}
The data that support the findings of this study are available within the article and its Supporting Information files. Raw datasets generated during the density functional theory calculations have been deposited in the Zenodo repository at DOI: \href{https://doi.org/10.5281/zenodo.21923014}{10.5281/zenodo.21923014}.

\bmhead{Code Availability}
First-principles calculations for the isolated molecular building blocks were performed using the open-source code \texttt{Octopus}\cite{TancogneDejean2020}. Hybrid interface calculations were modeled using the proprietary \texttt{VASP} software~\cite{kres-fuer96prb} under a valid license. The custom Python-based post-processing scripts used to reconstruct the molecular
hyperpolarizability tensors, perform tensor rotations, evaluate the model
substrate response, and generate the SHG screening figures are archived together
with the Zenodo dataset at DOI: \href{https://doi.org/10.5281/zenodo.21923014}{10.5281/zenodo.21923014}. A documented and generalized software package implementing this workflow will
be described in a forthcoming methods paper.

%The custom Python-based codes to perform the optical screening framework has been deposited on GitHub [INSERT LINK HERE].

\bmhead{Acknowledgements}
The authors thank Sumanti Patra for useful discussions.

\bmhead{Author Contributions}
M.G. developed the model, conducted all first-principles calculations on isolated molecules, ran the screening of the polarizability tensors, curated the data, and wrote the first version of the manuscript. M.S.R. conducted the fully atomistic first-principles analysis of selected hybrid interfaces. 
C.C. conceived the project, acquired the funding, validated the analysis, and finalized the manuscript. All authors reviewed and edited the final text.

\bmhead{Competing Interests}
The authors declare no competing financial or non-financial interests.

\bmhead{Ethics approval and consent to participate}
Not applicable.

\bmhead{Consent for publication}
Not applicable.

\bmhead{Materials availability}
Not applicable.

%%===========================================================================================%%
%% If you are submitting to one of the Nature Portfolio journals, using the eJP submission   %%
%% system, please include the references within the manuscript file itself. You may do this  %%
%% by copying the reference list from your .bbl file, paste it into the main manuscript .tex %%
%% file, and delete the associated \verb+\bibliography+ commands.                            %%
%%===========================================================================================%%
\bibliographystyle{naturemag}
%\bibliography{references}% common bib file
%% if required, the content of .bbl file can be included here once bbl is generated
%%\input sn-article.bbl

\end{document}